\documentclass[conference]{IEEEtran}
\usepackage{graphicx,amssymb,amsmath}
\usepackage[noadjust]{cite}
\usepackage{setspace}               
\usepackage{stfloats}
\usepackage{bbding}
\usepackage{amsthm}
\usepackage{amsmath}
\usepackage{flushend}
\usepackage{cases,subeqnarray}
\usepackage{bm,multirow,bigstrut}
\usepackage{textcomp}
\usepackage{latexsym,bm}
\usepackage{booktabs,changebar}
\usepackage{xcolor}
\usepackage{mathtools}
\usepackage{dsfont}
\usepackage{extarrows}
\usepackage{mathrsfs}
\usepackage{subfigure}
\usepackage{pifont}

\usepackage{cite}
\usepackage{bm}
\usepackage{cleveref}
\usepackage{multicol}       
\usepackage{multirow}       
\usepackage{array}          
\usepackage{colortbl}
\usepackage{makecell}
\definecolor{crimson}{RGB}{192,0,0}         
\definecolor{navy}{RGB}{47,85,151}         

\makeatletter                               
\newif\if@restonecol
\makeatother

\makeatletter
\newif\if@restonecol
\makeatother

\usepackage[linesnumbered,ruled,vlined]{algorithm2e}
\usepackage{algpseudocode}
\usepackage{amsmath}
\usepackage{listings}                       
\usepackage[framed,numbered,autolinebreaks,useliterate]{mcode}
\theoremstyle{plain}

\theoremstyle{plain}

\usepackage[ruled,vlined]{algorithm2e}
\SetAlgoSkip{}           
\SetAlgoInsideSkip{}

\IEEEoverridecommandlockouts

\begin{document}

\title{Harnessing the Freedom of Non-Uniformity in Monostatic ISAC with Antenna Flexibility \vspace{-0.3cm}}

\author{Zhe Wang, Mahmoud Zaher, Vitaly Petrov, and Emil Bj{\"o}rnson\\
{\small KTH Royal Institute of Technology and Digital Futures, Stockholm, Sweden.}\\
{\small Email: \{zhewang2,mahmoudz,vitalyp,emilbjo\}@kth.se}\vspace{-0.3cm}
\thanks{This work was supported by the research center Digital Futures in Sweden, and by the Grant 2022-04222 from the Swedish Research Council.}}

\maketitle


\begin{abstract}
This paper studies flexible non-uniform array design for monostatic integrated sensing and communication (ISAC) systems. An antenna pool is considered at the base station, where each candidate antenna can be dynamically assigned to transmit, receive, or inactive modes, such that a non-uniform effective array is jointly constructed with the ISAC precoding design. We formulate a sum communication rate maximization problem by jointly optimizing the ISAC beamforming schemes and antenna-mode assignment under sensing, power, and antenna mode constraints. We develop an alternating-optimization-based solution framework mainly with the aid of weighted minimum mean square error, continuous relaxation-based penalty, and successive convex approximation. Numerical results show that the proposed non-uniform array achieves higher sum-rates than the uniform-array baselines, with particularly large gains when the number of activated antennas is small. Moreover, the proposed non-uniform array can achieve, and in some cases exceed, the performance of uniform array baselines with substantially fewer activated antennas, highlighting geometry-aware non-uniform array design as a compelling alternative to brute-force antenna scaling-based array design.

\end{abstract}

\IEEEpeerreviewmaketitle
\section{Introduction}
\bstctlcite{BSTcontrol}
Massive multiple-input multiple-output (MIMO) arrays in 5G are built on the philosophy of antenna abundance, where performance is improved by deploying many antennas in uniform, half-wavelength-spaced structures \cite{ZheSurvey,5595728,6736761}. While this design principle has been highly successful, it largely treats array geometry as fixed and provides limited spatial flexibility \cite{bjornson2026antenna}. However, as wireless systems evolve toward higher carrier frequencies under tighter cost and energy constraints, simply increasing the antenna count is no longer a sustainable solution. This has motivated a shift in array design from antenna abundance to antenna flexibility, where performance improvements are achieved through flexible geometry reconfiguration rather than enlarging fixed uniform structures \cite{10906511,bjornson2026antenna,wang2025flexible}. In this context, non-uniform arrays have emerged as a promising design paradigm. Their realizations can take many forms, including movable antenna arrays \cite{10906511}, site-specific pre-optimized arrays \cite{irshad2025pre}, predefined structural layout-driven arrays \cite{9465145}, and modular array architectures \cite{10545312}. Despite their different realizations, these approaches share the same philosophy: array geometry is exploited as a design dimension rather than treated as a fixed hardware structure, which is particularly valuable when only a limited number of antennas can be utilized.

This perspective is especially relevant to monostatic integrated sensing and communication (ISAC) systems, where the same antenna platform must simultaneously support multi-user transmission, target illumination, and echo reception \cite{9737357,11159291,TWC23CKJ1}. A common implementation is to employ two separate arrays, one for transmit (Tx) operation and the other for receive (Rx) operation \cite{10158711,9724187,10159012}. While such a configuration is convenient, it fixes both the array geometry and the Tx/Rx partition a priori, so that the effective Tx and Rx apertures cannot adapt to the current user and target configurations. More recently, dynamic Tx/Rx partitioning of a shared array has also been considered for monostatic ISAC, jointly optimized with beamforming \cite{10810291}. \emph{However, existing array designs either rely on predetermined disjoint Tx/Rx arrays or, at most, optimize Tx/Rx partitioning within a shared array, without exploiting the additional flexibility offered by antenna inactivation when shaping effective non-uniform array geometries via a common antenna platform.} Consequently, it remains unclear whether monostatic ISAC can benefit from a more general antenna-role assignment mechanism that not only determines Tx/Rx functions, but also forms effective non-uniform arrays according to dynamic user and target configurations. This issue becomes particularly important when only a limited number of antennas are activated to save power, since performance then depends more critically on the geometry and role of each antenna.

Motivated by the above observations, we study non-uniform array design in ISAC systems. The main contributions are summarized as follows. First, we introduce a candidate-array-based non-uniform array design for monostatic ISAC, where an antenna candidate pool is given, and a non-uniform effective array is constructed through transmit/receive/inactivation antenna-mode assignment. This provides a flexible and practically tractable way to realize non-uniform arrays in monostatic ISAC. Then, we formulate a joint ISAC design problem that couples non-uniform array construction with beamforming and antenna-mode optimization. A sum communication rate maximization problem is formulated under sensing, power, and antenna mode constraints, where the ISAC beamforming schemes and antenna assignment strategies are jointly optimized. Finally, we develop an alternating optimization-based solution by combining the weighted minimum mean square error, continuous relaxation-based penalty, and successive convex approximation (SCA), and progressive antenna hardening strategy in a novel manner.

\section{System Model and Problem Formulation}\label{sec:system}

We study a downlink ISAC system, where a base station (BS) provides communication service to $K$ single-antenna user equipments (UEs) and performs monostatic target position detection for a sensing target. The BS is equipped with an array of $N$ candidate antennas with arbitrary shape, which can be optimized to be dynamically switched between three mutually exclusive antenna-modes: transmitting mode, receiving mode, and inactive mode. We define the array scheduling vectors $\mathbf{a}_T=\left[ a_{1,T},a_{2,T},\dots ,a_{N,T} \right] ^T\in \left\{ 0,1 \right\} ^N$ and $\mathbf{a}_R=\left[ a_{1,R},a_{2,R},\dots ,a_{N,R} \right] ^T\in \left\{ 0,1 \right\} ^N$ to describe the transmitting and receiving modes for the antenna array, respectively, where $a_{n,T}=1$ and $a_{n,R}=1$ denote that the $n$-th antenna is scheduled to be a transmit antenna and a receive antenna, respectively. Notably, each antenna is assigned to exactly one of the three antenna roles, and thus, we have $a_{n,T}+a_{n,R}\leqslant 1\,\,\, \forall n$. 

We consider a transmission block of $B$ channel uses for both communication and sensing, with static channel realizations. We utilize the deterministic sensing signal sequence $\mathbf{s}_0=[ s_{0,1},\dots ,s_{0,B} ] ^T\in \mathbb{C} ^B$ for the target sensing with $B$ being the sequence length, $| s_{0,i} |^2=1$ for $i\in \{ 1,\dots ,B \} $ and  $\| \mathbf{s}_0 \| ^2=B$. We formulate the downlink ISAC signal in the $i$-th channel use as 
\vspace{-0.55em}\begin{equation}
\mathbf{x}_i=\mathbf{A}_T\left( \sum_{k=1}^K{\mathbf{v}_ks_{k,i}}+\mathbf{v}_0s_{0,i} \right) \in \mathbb{C} ^N,
\vspace{-0.3em}\end{equation}
where $\mathbf{A}_T=\mathrm{diag}\left( \mathbf{a}_T \right) \in \mathbb{R} ^{N\times N}$ is the transmit antenna assignment matrix, $\mathbf{v}_k\in \mathbb{C} ^N$ is the precoding vector for UE $k$, $s_{k,i}\sim \mathcal{N} _{\mathbb{C}}(0,1)$ is the intended signal for UE $k$, $s_{0,i}\in \mathbb{C}$ is the $i$-th sensing signal in $\mathbf{s}_0$, and $\mathbf{v}_0\in \mathbb{C} ^N$ denotes the sensing precoding vector. The covariance matrix for $\mathbf{x}_i$ can be constructed as $\mathbf{R}_x=\mathbb{E} \{ \mathbf{xx}^H \} =\mathbf{A}_T( \sum_{k=1}^K{\mathbf{v}_k\mathbf{v}_{k}^{H}}+\mathbf{v}_0\mathbf{v}_{0}^{H} ) \mathbf{A}_T$, where the transmit power is constrained by $\mathrm{tr}\left( \mathbf{R}_x \right) \leqslant P_{\max}$ with $P_{\max}$ being the maximum transmit power.

For the downlink communication, the received downlink signal at UE~$k$ in the $i$-th channel use is 
\vspace{-0.55em}\begin{equation}
y_{k,i}=\mathbf{h}_{k}^{H}\mathbf{A}_T\mathbf{v}_ks_{k,i}+\sum_{l\ne k}^K{\mathbf{h}_{k}^{H}\mathbf{A}_T\mathbf{v}_ls_{l,i}}+\mathbf{h}_{k}^{H}\mathbf{A}_T\mathbf{v}_0s_{0,i}+n_{k,i},
\vspace{-0.3em}\end{equation} 
where $n_{k,i}\sim \mathcal{N} _{\mathbb{C}}(0,\sigma _{k}^{2})$ is the additive noise with $\sigma _{k}^{2}$ being the noise power. The achievable communication rate for UE $k$ is $R_k=\log _2\left( 1+\mathrm{SINR}_k \right) $ with the signal-to-interference-plus-noise ratio (SINR) being
\vspace{-0.55em}\begin{equation}
\mathrm{SINR}_k=\frac{\left| \mathbf{h}_{k}^{H}\mathbf{A}_T\mathbf{v}_k \right|^2}{\sum_{l\ne k}^K{\left| \mathbf{h}_{k}^{H}\mathbf{A}_T\mathbf{v}_l \right|^2}+\left| \mathbf{h}_{k}^{H}\mathbf{A}_T\mathbf{v}_0 \right|^2+\sigma _{k}^{2}}.
\vspace{-0.3em}\end{equation}

For the sensing operation, the BS processes the received echo signal within a block to perform target detection. This received signal at the BS in the $i$-th channel use is 
\begin{align} \nonumber
\mathbf{y}_{i}\!&=\!\!\mathbf{A}_R\left( \alpha _0\mathbf{g}_0\mathbf{g}_{0}^{T}\mathbf{x}_i+\mathbf{H}_{\mathrm{SI}}\mathbf{x}_i+\mathbf{n}_{BS,i}\right )\!\! =\!\!\mathbf{A}_R\alpha _0\mathbf{g}_0\mathbf{g}_{0}^{T}\mathbf{A}_T\mathbf{v}_0{s}_{0,i}\\
&+\mathbf{A}_R \left( \alpha _0\mathbf{g}_0\mathbf{g}_{0}^{T}\mathbf{A}_T \sum_{k=1}^K{\mathbf{v}_ks_{k,i}}  +\mathbf{H}_{\mathrm{SI}}\mathbf{x}_i+\mathbf{n}_{BS,i} \right),
\end{align}
where $\mathbf{A}_R=\mathrm{diag}\left( \mathbf{a}_R \right) \in \mathbb{R} ^{N\times N}$ represents the receive antenna assignment matrix, $\alpha _0\sim \mathcal{N} _{\mathbb{C}}\left( 0,\sigma _{0}^{2} \right) $ denotes the unknown radar cross section (RCS) of the target, $\mathbf{H}_{\mathrm{SI}}\in \mathbb{C} ^{N\times N}$ is the self-interference (SI) channel between the transmit and receive antennas at the BS, $\mathbf{g}_0\in \mathbb{C} ^N$ is the channel between the BS and target, and $\mathbf{n}_{BS,i}\sim \mathcal{N} _{\mathbb{C}}( \mathbf{0},\sigma _{r}^{2}\mathbf{I}_N ) $ is the receiver noise at the BS with $\sigma _{r}^{2}$ being the noise power. We gather the received signals within the sensing block in $\mathbf{Y}=[ \mathbf{y}_1,\dots ,\mathbf{y}_B ] \in \mathbb{C} ^{N\times B}$. To suppress interference, we project $\mathbf{Y}$ onto the sensing signal $\mathbf{s}_0$ and apply a receive combining vector $\mathbf{u}\in \mathbb{C} ^N$, which leads to $\mathbf{u}^H{\mathbf{Ys}_{0}^{*}}$. The resulting sensing SINR for target detection is

%

\vspace{-0.55em}\begin{equation}\label{sensing_SINR}
\begin{aligned}
\mathrm{SINR}_0\!\!=\!\!\frac{B\sigma _{0}^{2}\left| \mathbf{u}^H\mathbf{A}_R\mathbf{G}_0\mathbf{A}_T\mathbf{v}_0 \right|^2}{\mathbf{u}^H\mathbf{A}_R\left( \sigma _{0}^{2}\mathbf{G}_0\mathbf{R}_c\mathbf{G}_{0}^{H}+\mathbf{H}_{\mathrm{SI}}\mathbf{R}_x\mathbf{H}_{\mathrm{SI}}^{H}+\sigma _{r}^{2}\mathbf{I}_N \right) \mathbf{A}_R\mathbf{u}},
\end{aligned}
\vspace{-0.3em}\end{equation}
where $\mathbf{G}_0=\mathbf{g}_0\mathbf{g}_{0}^{T}$, $\mathbf{R}_0=\mathbf{A}_T\mathbf{v}_0\mathbf{v}_0^H\mathbf{A}_{T}$, and $\mathbf{R}_c=\mathbf{A}_T( \sum_{k=1}^K{\mathbf{v}_k\mathbf{v}_{k}^{H}} ) \mathbf{A}_{T}$. Note that a higher sensing SINR generally implies better sensing quality, as it improves the reliability of the target echo relative to interference and noise, thereby leading to more accurate target detection, localization, or tracking. Hence, we use the sensing SINR to measure the sensing performance in the following.

We study a sum communication rate maximization problem under a sensing performance constraint, where the beamforming schemes and antenna assignment strategies are jointly optimized. Specifically, we aim to jointly optimize the transmit precoding schemes $\left\{\mathbf{v}_k \right\} _{k=1}^{K}$, sensing precoding scheme $\mathbf{v}_0$, sensing receive combining scheme $\mathbf{u}$, and antenna assignment strategies $\mathbf{a}_T$, $\mathbf{a}_R$. Let $\mathcal{A} =\{ \{ \mathbf{v}_k \} _{k=1}^{K},\mathbf{v}_0,\mathbf{u},\mathbf{a}_T,\mathbf{a}_R \} $ denote the set of these optimization variables. Then, we construct the sum-rate maximization problem as 
\vspace{-0.55em}\begin{equation}\label{sum_rate_maximization}
\begin{aligned}
&\underset{\mathcal{A}}{\max}\sum_{k=1}^K{R_k}\\
&\,\mathrm{s}.\mathrm{t}. \, \,\mathrm{SINR}_0\geqslant \gamma _0, \mathrm{tr}\left( \mathbf{R}_x \right) \leqslant P_{\max}, \mathbf{a}_T, \mathbf{a}_R\in \left\{ 0,1 \right\} ^N,\\
&\,\,\,\,\, \, \,\,\,\,\,   a_{n,T}+a_{n,R}\leqslant 1, \sum_{n=1}^N{\left( a_{n,T}+a_{n,R} \right)}\leqslant N_{\mathrm{act}}, \forall n,\\
\end{aligned}
\vspace{-0.3em}\end{equation}
where $\gamma_0$ is the sensing SINR threshold and $N_\mathrm{act}\leqslant N$ denotes the maximum number of activated antennas among all $N$ candidate antennas (e.g., selected to limit energy consumption).

\section{Joint ISAC Design for Sum-Rate Maximization}\label{sec:sumrate}

The problem in \eqref{sum_rate_maximization} is non-convex. Hence, in this section, we solve by alternating optimization (AO), where we iteratively optimize one variable set while fixing the remaining variables.

\subsection{Solution to Receive Combining Scheme}


We start with the optimization of $\mathbf{u}$. We can easily express $\mathrm{SINR}_0$ in \eqref{sensing_SINR} as a Rayleigh quotient with respect to $\mathbf{u}$, and then the optimal $\mathbf{u}$ maximizing $\mathrm{SINR}_0$ can be derived based on the Rayleigh quotient maximization lemma \cite{8187178} as
\vspace{-0.55em}\begin{equation}\label{opt_u}
\mathbf{u}^{\star}\!\!=\!\!\left[ \mathbf{A}_R\left( \sigma _{0}^{2}\mathbf{G}_0\mathbf{R}_c\mathbf{G}_{0}^{H}+\mathbf{H}_{\mathrm{SI}}\mathbf{R}_x\mathbf{H}_{\mathrm{SI}}^{H}+\sigma _{r}^{2}\mathbf{I}_N \right) \mathbf{A}_R \right] ^{-1}\!\!\mathbf{g}_{0,R},
\vspace{-0.3em}\end{equation}
where $\mathbf{g}_{0,R}=\mathbf{A}_R\mathbf{g}_0$ denotes the effective receiving array channel response for sensing for current receive antenna set. 

\subsection{Solution to Transmit Precoding Scheme}\label{Sec_solution_beamforming}
\vspace{-0.1cm}
In this part, we focus on the optimization of the precoding vectors $\{\mathbf{v}_k\}_{k=1}^K$, $\mathbf{v}_0$. We utilize the weighted minimum mean square error (WMMSE) approach from \cite{5756489} to transform the sum-rate maximization in \eqref{sum_rate_maximization} into a weighted sum mean square error (MSE) minimization problem. More specifically, the decoding MSE of $s_{k,i}$ can be formulated as 
$e_k=\left| c_k \right|^2( \sum_{l=1}^K{\left| \mathbf{h}_{k}^{H}\mathbf{A}_T\mathbf{v}_l \right|^2}+\left| \mathbf{h}_{k}^{H}\mathbf{A}_T\mathbf{v}_0 \right|^2+\sigma _{k}^{2} ) -2\Re \{ c_{k}^{*}\mathbf{h}_{k}^{H}\mathbf{A}_T\mathbf{v}_k \} +1$, where $ c_{k}$ denotes an auxiliary scalar combining coefficient for the decoding. The optimal $ c_{k}$, which minimizes the decoding MSE, is 
$c_{k}^{\star}\!=\!\left( \sum_{l=1}^K{\left| \mathbf{h}_{k}^{H}\mathbf{A}_T\mathbf{v}_l \right|^2}+\left| \mathbf{h}_{k}^{H}\mathbf{A}_T\mathbf{v}_0 \right|^2+\sigma _{k}^{2} \right) ^{-1}\!\!\!\mathbf{h}_{k}^{H}\mathbf{A}_T\mathbf{v}_k$
Then, by taking the classical WMMSE approach, we define the scalar weights $w_k>0$ and the sum-rate $\sum_{k=1}^K{R_k}$ maximization problem is equivalent to the scalar-weighted sum-MSE $\sum_{k=1}^K{\left( w_ke_k-\log w_k \right)}$ minimization problem, where $w_k$ is updated as $w_k=1/e_k$ \cite{5756489}. We also utilize this optimization objective during the following optimization of antenna assignment strategies. Thus, when the other variables are fixed, the sum-rate maximization problem can be equivalently constructed as (we omit the unaffected constant parts)
\vspace{-0.55em}\begin{equation}\label{sum_mse_minimization}
\begin{aligned}
&\underset{\left\{ \mathbf{v}_k \right\} ,\mathbf{v}_0}{\min}\sum_{k=1}^K{\mathbf{v}_{k}^{H}\mathbf{\Upsilon v}_k}+\mathbf{v}_{0}^{H}\mathbf{\Upsilon v}_0 -2\sum_{k=1}^K{\Re \left\{ \mathbf{b}_{k}^{H}\mathbf{v}_k \right\}}\\
&\,\,\,\,\,\,\,\,\mathrm{s}.\mathrm{t}.\,\,\, \mathrm{SINR}_0\geqslant \gamma _0, \mathrm{tr}\left( \mathbf{R}_x \right) \leqslant P_{\max},
\end{aligned}
\vspace{-0.3em}\end{equation}
with $\mathbf{\Upsilon }\!=\sum_{k=1}^K{w_k\left| c_k \right|^2\mathbf{A}_T\mathbf{h}_k\mathbf{h}_{k}^{H}\mathbf{A}_T}$ and $ \mathbf{b}_k=w_kc_k\mathbf{A}_T\mathbf{h}_k$. 

We first fix $\mathbf{v}_0$ and optimize $\{\mathbf{v}_k\}_{k=1}^K$. The power constraint $\mathrm{tr}\left( \mathbf{R}_x \right) \leqslant P_{\max}$ can be denoted as $ \sum_{k=1}^K{\| \mathbf{A}_T\mathbf{v}_k \| _{2}^{2}}\leqslant P_{\max}- \| \mathbf{A}_T\mathbf{v}_0 \| _{2}^{2}  $. The sensing constraint $\mathrm{SINR}_0\geqslant \gamma _0$ can be transformed to 
\vspace{-0.55em}\begin{equation}\label{v_k_sensing_constraint}
\begin{aligned}
&{B\sigma _{0}^{2}}\mathbf{v}_{0}^{H}\mathbf{Q}_{G}\mathbf{v}_0-\gamma _0\mathbf{v}_{0}^{H}\mathbf{Q}_{\mathrm{SI}}\mathbf{v}_0-\gamma _0\sigma _{r}^{2}\mathbf{u}^H\mathbf{A}_R\mathbf{A}_R\mathbf{u}\\
&\geqslant\gamma _0\sigma _{0}^{2}\sum_{k=1}^K{\mathbf{v}_{k}^{H}\mathbf{Q}_G\mathbf{v}_k}+\gamma _0\sum_{k=1}^K{\mathbf{v}_{k}^{H}\mathbf{Q}_{\mathrm{SI}}\mathbf{v}_k},
\end{aligned}
\vspace{-0.3em}\end{equation}
where $\mathbf{Q}_G=\mathbf{A}_T\mathbf{G}_{0}^{H}\mathbf{A}_R\mathbf{uu}^H\mathbf{A}_R\mathbf{G}_0\mathbf{A}_T\succeq 0$ and $\mathbf{Q}_{\mathrm{SI}}=\mathbf{A}_T\mathbf{H}_{\mathrm{SI}}^{H}\mathbf{A}_R\mathbf{uu}^H\mathbf{A}_R\mathbf{H}_{\mathrm{SI}}\mathbf{A}_T\succeq 0$.
Thus, the optimization problem for $\left\{ \mathbf{v}_k \right\}$ is formulated as
\vspace{-0.55em}\begin{equation}\label{update_v_k}
\begin{aligned}
&\underset{\left\{ \mathbf{v}_k \right\}}{\min}\,\sum_{k=1}^K{\mathbf{v}_{k}^{H}\mathbf{\Upsilon v}_k}-2\sum_{k=1}^K{\Re \left\{ \mathbf{b}_{k}^{H}\mathbf{v}_k \right\}}\\
&\,\mathrm{s}.\mathrm{t}. \,\,\,\eqref{v_k_sensing_constraint}, \,\, \sum\nolimits_{k=1}^K{\left\| \mathbf{A}_T\mathbf{v}_k \right\| _{2}^{2}}\leqslant P_{\max}-\left\| \mathbf{A}_T\mathbf{v}_0 \right\| _{2}^{2},
\end{aligned}
\vspace{-0.3em}\end{equation}
which is a standard quadratically constrained quadratic program (QCQP), and can be efficiently solved by general-purpose convex solvers.

For the optimization of $\mathbf{v}_0$, we apply the SCA to $\mathbf{v}_{0}^{H}\mathbf{Q}_G\mathbf{v}_0$ in \eqref{v_k_sensing_constraint} at the result in the $t$-th iteration $\mathbf{v}_{0}^{\left( t \right)}$ and derive the approximated sensing constraint in iteration $(t+1)$ as 
\vspace{-0.55em}\begin{equation}\label{v_0_sensing_constraint}
\begin{aligned}
&B\sigma _{0}^{2}( 2\Re \{ ( \mathbf{v}_{0}^{( t )} ) ^H\mathbf{Q}_G\mathbf{v}_0 \} -( \mathbf{v}_{0}^{( t )} ) ^H\mathbf{Q}_G\mathbf{v}_{0}^{( t )} ) -\gamma _0\mathbf{v}_{0}^{H}\mathbf{Q}_{\mathrm{SI}}\mathbf{v}_0\\
&\geqslant \gamma _0\sigma _{0}^{2}\sum_{k=1}^K{\mathbf{v}_{k}^{H}\mathbf{Q}_G\mathbf{v}_k}+\gamma _0\sum_{k=1}^K{\mathbf{v}_{k}^{H}\mathbf{Q}_{\mathrm{SI}}\mathbf{v}_k}+\gamma _0\sigma _{r}^{2}\mathbf{u}^H\mathbf{A}_R\mathbf{u}.
\end{aligned}
\vspace{-0.3em}\end{equation}
Thus, we construct the optimization problem for $\mathbf{v}_0$ as
\vspace{-0.55em}\begin{equation}\label{update_V_0}
\begin{aligned}
&\underset{\mathbf{v}_0}{\min}\,\,\mathbf{v}_{0}^{H}\mathbf{\Upsilon v}_0 \,\,\\
&\,\,\mathrm{s}.\mathrm{t}. \,\,\, \eqref{v_0_sensing_constraint},\,\,\left\| \mathbf{A}_T\mathbf{v}_0 \right\| _{2}^{2} \leqslant P_{\max}-\sum\nolimits_{k=1}^K{\left\| \mathbf{A}_T\mathbf{v}_k \right\| _{2}^{2}},
\end{aligned}
\vspace{-0.3em}\end{equation}
which can be solved by general-purpose convex solvers.

\subsection{Solution to Antenna Assignment Strategy}
In this part, we focus on the optimization of antenna assignment strategies $\mathbf{a}_T$ and $\mathbf{a}_R$. We apply the continuous relaxation operation for $a_{n,T},a_{n,R}\in \left\{ 0,1 \right\} $ to $a_{n,T},a_{n,R}\in [0,1]$ while maintaining the constraints $a_{n,T}+a_{n,R}\leqslant 1 $, $\sum_{n=1}^N{a_{n,T}}\geqslant N_{T,\min}$, and $\sum_{n=1}^N{\left( a_{n,T}+a_{n,R} \right)}\leqslant N_{\mathrm{act}}$. To facilitate  $a_{n,T}$ and $a_{n,R}$ to be binary, we perform penalty-driven optimization, where the penalty function is defined as $\Phi \left( \mathbf{a}_T,\mathbf{a}_R \right) =\sum_{n=1}^N{( a_{n,T}-a_{n,T}^{2} )}+\sum_{n=1}^N{( a_{n,R}-a_{n,R}^{2} )}$. Note that $\Phi \left( \mathbf{a}_T,\mathbf{a}_R \right) \geqslant 0$  when $0\leqslant a_{n,T}\leqslant 1$ and $ 0\leqslant a_{n,R}\leqslant 1$ and only $\Phi \left( \mathbf{a}_T,\mathbf{a}_R \right) =0$ when $a_{n,T},a_{n,R}\in \{ 0,1 \} $, $\forall n$. With the penalty function involved, the objective is formulated as $\underset{\mathbf{a}_T,\mathbf{a}_R}{\min}\sum_{k=1}^K{w_ke_k}+\mu \Phi ( \mathbf{a}_T,\mathbf{a}_R ) $ with $\mu>0$ being the penalty parameter.

We first focus on the optimization of $\mathbf{a}_T$ while fixing $\mathbf{a}_R$ and omitting $\mathbf{a}_R$-related terms in the objective. 
We apply SCA method on $-a^2$ and derive the approximated penalty function for the transmitting mode as $\Phi ^{( t+1 )}( \mathbf{a}_T ) =\sum_{n=1}^N{[ ( 1-2a_{n,T}^{( t )} ) a_{n,T}+( a_{n,T}^{( t )} ) ^2 ]}$. Then, we formulate $e_k$ as 
\vspace{-0.55em}\begin{equation}
e_k=\mathbf{a}_{T}^{T}\mathbf{Q}_k\mathbf{a}_T-2\bar{\mathbf{q}}_{k}^{T}\mathbf{a}_T+\mathrm{const}
\vspace{-0.3em}\end{equation}
with the aid of some matrix transformations, where $\mathbf{Q}_k=| c_k |^2( \sum_{l=1}^K{\mathbf{Q}_{kl}}+\mathbf{Q}_{k0} )\in \mathbb{R} ^{N\times N} $, $\mathbf{Q}_{kl}=\Re \{ \mathbf{q}_{kl}\mathbf{q}_{kl}^{H} \} \succeq 0\in \mathbb{R} ^{N\times N} $, $\mathbf{q}_{kl}=\mathbf{h}_{k}^{*}\odot \mathbf{v}_l\in \mathbb{C} ^N$, $\mathbf{Q}_{k0}=\Re \left\{ \mathbf{q}_{k0}\mathbf{q}_{k0}^{H} \right\} \succeq 0$, $\mathbf{q}_{k0}=\mathbf{h}_{k}^{*}\odot \mathbf{v}_0\in \mathbb{C} ^N$, $\bar{\mathbf{q}}_k=\Re \{ c_{k}^{*}\mathbf{q}_k \} \in \mathbb{R} ^N$, and $\mathrm{const}$ is a constant which is not related to $\mathbf{a}_T$. Note that the real-valued operator $\Re\{\cdot\}$ is implemented with the aid of the real-valued feature for $\mathbf{a}_T$, which is handy for the implementation of the following SCA technique. Furthermore, we have $\sum_{k=1}^K{w_ke_k}=\mathbf{a}_{T}^{T}\mathbf{Qa}_T-2\bar{\mathbf{q}}^T\mathbf{a}_T+\mathrm{const}$ with $\mathbf{Q}=\sum_{k=1}^K{w_k\mathbf{Q}_k}$ and $\bar{\mathbf{q}}=\sum_{k=1}^K{w_k}\bar{\mathbf{q}}_k$. For the power constraint, we have $\mathbf{a}_{T}^{T}\mathbf{P}_{\mathrm{pow}}\mathbf{a}_T\leqslant P_{\max}$ with
$\mathbf{P}_{\mathrm{pow}}=\Re \{ \mathrm{diag}( \sum_{k=1}^K{| \mathbf{v}_k |^2}+| \mathbf{v}_0 |^2 ) \}\in \mathbb{R} ^{N\times N} $. As for the sensing constraint, we can first construct it as $\sigma _{0}^{2}\mathrm{tr}\left( \mathbf{A}_T \mathbf{v}_0 \mathbf{v}_0^H\mathbf{A}_T\mathbf{M}_G \right) -\gamma _0\sigma _{0}^{2}\mathrm{tr}\left( \mathbf{A}_T\overline{\mathbf{R}}_c\mathbf{A}_T\mathbf{M}_G \right) -\gamma _0\mathrm{tr}\left( \mathbf{A}_T\overline{\mathbf{R}}_x\mathbf{A}_T\mathbf{M}_{\mathrm{SI}} \right) \geqslant \gamma _0\sigma _{r}^{2}\mathbf{u}^H\mathbf{A}_R\mathbf{A}_R\mathbf{u}$, where $\mathbf{M}_G=\mathbf{G}_{0}^{H}\mathbf{U}_R\mathbf{G}_0\succeq 0\in \mathbb{C} ^{N\times N}$  $\mathbf{U}_R=\mathbf{A}_R\mathbf{uu}^H\mathbf{A}_R\in \mathbb{C} ^{N\times N}$, and $\mathbf{M}_{\mathrm{SI}}=\mathbf{H}_{\mathrm{SI}}^{H}\mathbf{U}_R\mathbf{H}_{\mathrm{SI}}\succeq 0\in \mathbb{C} ^{N\times N}$. Then, we have $\sigma _{0}^{2}\mathbf{a}_{T}^{T}\mathbf{B}_0\mathbf{a}_T\geqslant \gamma _0\sigma _{0}^{2}\mathbf{a}_{T}^{T}\mathbf{B}_c\mathbf{a}_T+\gamma _0\mathbf{a}_{T}^{T}\mathbf{B}_{\mathrm{SI}}\mathbf{a}_T+\gamma _0\sigma _{r}^{2}\mathbf{u}^H\mathbf{A}_R\mathbf{A}_R\mathbf{u}$, where $\mathbf{B}_0=\Re \{  (\mathbf{v}_0 \mathbf{v}_0^H)\odot \mathbf{M}_{G}^{T} \}\succeq 0 \in \mathbb{R} ^{N\times N}$, $\mathbf{B}_c=\Re \{ \overline{\mathbf{R}}_c\odot \mathbf{M}_{G}^{T} \}\succeq 0 \in \mathbb{R} ^{N\times N}$ $\mathbf{B}_{\mathrm{SI}}=\Re \{ \overline{\mathbf{R}}_x\odot \mathbf{M}_{\mathrm{SI}}^{T} \} \succeq 0 \in \mathbb{R} ^{N\times N}$. We implement the SCA method for $\sigma _{0}^{2}\mathbf{a}_{T}^{T}\mathbf{B}_0\mathbf{a}_T$ at the result in the $t$-th iteration $\mathbf{a}_{T}^{\left( t \right)}$ and derive the approximated sensing constraint in iteration $t+1$ as 
\vspace{-0.55em}\begin{equation}\label{a_t_sensing_constraint}
\begin{aligned}
&\gamma _0\sigma _{0}^{2}\mathbf{a}_{T}^{T}\mathbf{B}_c\mathbf{a}_T+\gamma _0\mathbf{a}_{T}^{T}\mathbf{B}_{\mathrm{SI}}\mathbf{a}_T\leqslant \sigma _{0}^{2}( \mathbf{a}_{T}^{( t )} ) ^T\mathbf{B}_0\mathbf{a}_{T}^{( t )}\\
&+2\sigma _{0}^{2}( \mathbf{a}_{T}^{( t )} ) ^T\mathbf{B}_0( \mathbf{a}_T-\mathbf{a}_{T}^{( t )} ) -\gamma _0\sigma _{r}^{2}\mathbf{u}^H\mathbf{A}_R\mathbf{A}_R\mathbf{u}.
\end{aligned}
\vspace{-0.3em}\end{equation}
Thus, we formulate the optimization problem for $\mathbf{a}_{T}$ in iteration $t+1$ as 
\vspace{-0.55em}\begin{equation}\label{update_a_t}
\begin{aligned}
&\underset{\mathbf{a}_T}{\min}\sum_{k=1}^K{w_ke_k}+\mu \Phi ^{\left( t+1 \right)}\left( \mathbf{a}_T \right) 
\\
&\,\mathrm{s}.\mathrm{t}.\,\,\, 0\leqslant a_{n,T}\leqslant 1, a_{n,T}+a_{n,R}\leqslant 1,\, \forall n, 
\\
&\,\,\,\,\,\,\,\,\,\,    \sum_{n=1}^N{\left( a_{n,T}+a_{n,R} \right)}\leqslant N_{\mathrm{act}}, \mathbf{a}_{T}^{T}\mathbf{P}_{\mathrm{pow}}\mathbf{a}_T\leqslant P_{\max},\,\,\eqref{a_t_sensing_constraint}
\end{aligned}
\vspace{-0.3em}\end{equation}
which can be solved by general-purpose convex solvers.

\begin{algorithm}[t!]
\caption{Progressive antenna hardening procedure}
\label{algo_hardening}

\KwIn{$\mathbf{a}_T,\mathbf{a}_R$, hardening state $\mathcal{F}=\{\mathcal{F}_{\mathrm{Tx}},\mathcal{F}_{\mathrm{Rx}},\mathcal{F}_{\mathrm{Off}}\}$, thresholds $\tau_{\{T,R\}},\tau_0,\tau_d$}

\KwOut{Hardened vectors $\mathbf{a}_T,\mathbf{a}_R$, hardening state $\mathcal{F}$}

\If{hardening is activated}
{\For{$n=1,\ldots,N$}
{\If{$(a_{T,n}\ge\tau_T$ and $a_{R,n}\le 1-\tau_T)$ or $(a_{T,n}-a_{R,n}\ge\tau_d$ and $a_{T,n}\ge 0.5)$}
{$f_{\mathrm{Tx},n}\leftarrow 1$;\\}

\If{$(a_{R,n}\ge\tau_R$ and $a_{T,n}\le 1-\tau_R)$ or $(a_{R,n}-a_{T,n}\ge\tau_d$ and $a_{R,n}\ge 0.5)$}
{$f_{\mathrm{Rx},n}\leftarrow 1$;\\}

\If{$a_{T,n}+a_{R,n}\le\tau_0$}
{$f_{\mathrm{Off},n}\leftarrow 1$;\\}

\If{$f_{\mathrm{Tx},n}=1$ or $f_{\mathrm{Rx},n}=1$}
{$f_{\mathrm{Off},n}\leftarrow 0$;\\}}

Update the cumulative hardening sets 
$\mathcal{F}_{\mathrm{Tx}} \leftarrow \mathcal{F}_{\mathrm{Tx}} \cup \{n: f_{\mathrm{Tx},n}=1\}$, $\mathcal{F}_{\mathrm{Rx}} \leftarrow \big(\mathcal{F}_{\mathrm{Rx}} \cup \{n: f_{\mathrm{Rx},n}=1\}\big)\setminus \mathcal{F}_{\mathrm{Tx}}$, $\mathcal{F}_{\mathrm{Off}} \leftarrow \big(\mathcal{F}_{\mathrm{Off}} \cup \{n: f_{\mathrm{Off},n}=1\}\big)\setminus \big(\mathcal{F}_{\mathrm{Tx}} \cup \mathcal{F}_{\mathrm{Rx}}\big)$;\\
Assign hard values to the frozen antennas as
$(a_{T,n},a_{R,n})=(1,0)$ for $n\in\mathcal{F}_{\mathrm{Tx}}$; $(a_{T,n},a_{R,n})=(0,1)$ for $n\in\mathcal{F}_{\mathrm{Rx}}$; $(a_{T,n},a_{R,n})=(0,0)$ for $n\in\mathcal{F}_{\mathrm{Off}}$;\\
}

Project $\mathbf{a}_T$ and $\mathbf{a}_R$ onto $[0,1]$ element-wisely;\\

\end{algorithm}

\begin{algorithm}[t!]
\caption{AO-Based Joint Beamforming and Antenna Assignment Algorithm}
\label{alg_overall_t}
\KwIn{$\{\mathbf{h}_k\}_{k=1}^K$, $\mathbf{g}_0$, $\mathbf{H}_{\mathrm{SI}}$, $N_{\mathrm{act}}$, $P_{\max}$, $\gamma_0$, maximum iteration number $T_{\max}$, tolerance $\epsilon$}
\KwOut{$\{\mathbf{v}_k\}_{k=1}^K$, $\mathbf{v}_0$, $\mathbf{u}$, $\mathbf{a}_{T}$, $\mathbf{a}_{R}$}

Initialize $\{\mathbf{v}_k^{(0)}\}_{k=1}^K$, $\mathbf{v}_0^{(0)}$, $\mathbf{a}_{T}^{(0)}$, and $\mathbf{a}_{R}^{(0)}$; \\

\For{$t=0,\ldots,T_{\max}-1$}
{
Update sensing combining $\mathbf{u}^{(t+1)}$ via \eqref{opt_u}; \\

Update WMMSE auxiliary variables $c_k^{(t+1)}$, $e_k^{(t+1)}$, and $w_k^{(t+1)}$, $\forall\,k$ based on Sec.~\ref{Sec_solution_beamforming}; \\

Update communication precoding $\{\mathbf{v}_k^{(t+1)}\}_{k=1}^K$ by solving \eqref{update_v_k} based on $\mathbf{v}_0^{(t)}$, $\mathbf{a}_{T}^{(t)}$, $\mathbf{a}_{R}^{(t)}$, and $\mathbf{u}^{(t+1)}$;\\

Update sensing precoding $\mathbf{v}_0^{(t+1)}$ by solving \eqref{update_V_0} based on $\{\mathbf{v}_k^{(t+1)}\}_{k=1}^K$, $\mathbf{a}_{T}^{(t)}$, $\mathbf{a}_{R}^{(t)}$, and $\mathbf{u}^{(t+1)}$;\\

Update transmit antenna assignment $\mathbf{a}_{T}^{(t+1)}$ by solving \eqref{update_a_t} based on $\mathbf{v}_0^{(t+1)}$, $\{\mathbf{v}_k^{(t+1)}\}_{k=1}^K$, $\mathbf{a}_{R}^{(t)}$, and $\mathbf{u}^{(t+1)}$;\\

Update receive antenna assignment $\mathbf{a}_{R}^{(t+1)}$ by solving \eqref{update_a_r} based on $\mathbf{v}_0^{(t+1)}$, $\{\mathbf{v}_k^{(t+1)}\!\}_{k=1}^K$, $\mathbf{a}_{T}^{(t+1)}$, and $\mathbf{u}^{(t+1)}$;\\

Apply adaptive antenna hardening to $\mathbf{a}_{T}^{(t+1)}$ and $\mathbf{a}_{R}^{(t+1)}$ via Algorithm~\ref{algo_hardening} if the triggering condition is satisfied, and update the hardening state $\mathcal{F}^{(t+1)}$; \\

Compute the objective $R^{(t+1)}=\sum_{k=1}^K{R_{k}^{(t+1)}}$; \\

\If{hardening stage is activated and 
$\delta_f^{(t+1)} \le \epsilon_{\mathrm{obj}}$ and
$\delta_T^{(t+1)} \le \epsilon_T$ and
$\delta_R^{(t+1)} \le \epsilon_R$}
{
\textbf{break};
}

}
\end{algorithm}

\addtolength{\topmargin}{0.05in}

For the optimization of $\mathbf{a}_R$, the transmit power is not related to $\mathbf{a}_R$, and thus, we focus on the transformation of the sensing constraint. Based on the SCA method, we represent the approximated sensing constraint based on $\mathbf{a}_{R}^{\left( t \right)}$ as 
\vspace{-0.55em}\begin{equation}\label{a_r_sensing_constraint}
\begin{aligned}
&\gamma _0\mathbf{a}_{R}^{T}\mathbf{C}_{c,SI}\mathbf{a}_R+\gamma _0\sigma _{r}^{2}\mathbf{a}_{R}^{T}\mathbf{D}_u\mathbf{a}_R\leqslant \\
&\sigma _{0}^{2}( \mathbf{a}_{R}^{( t )} ) ^T\mathbf{C}_0\mathbf{a}_{R}^{( t )}+2\sigma _{0}^{2}( \mathbf{a}_{R}^{\left( t \right)} ) ^T\mathbf{C}_0( \mathbf{a}_R-\mathbf{a}_{R}^{( t )} ),
\end{aligned}
\vspace{-0.3em}\end{equation}
where $\mathbf{C}_0=\Re \{ \mathrm{diag}( \mathbf{u} ) ^H\mathbf{N}_0\mathrm{diag}( \mathbf{u} ) \}\succeq 0 \in \mathbb{R} ^{N\times N}$, $\mathbf{C}_{c,SI}=\Re \{ \mathrm{diag}( \mathbf{u} ) ^H( \sigma _{0}^{2}\mathbf{N}_c+\mathbf{N}_{\mathrm{SI}} ) \mathrm{diag}( \mathbf{u} ) \}\succeq 0 \in \mathbb{R} ^{N\times N}$, and $\mathbf{D}_u=\mathrm{diag}( | u_1 |^2,\dots | u_N |^2 )\succeq 0 \in \mathbb{R} ^{N\times N} $. Note that $e_k$ is also not related to $\mathbf{a}_{R}$, and thus, we construct the optimization problem for $\mathbf{a}_{R}$ in iteration $t+1$ as 
\vspace{-0.55em}\begin{equation}\label{update_a_r}
\begin{aligned}
&\underset{\mathbf{a}_R}{\min}\,\,\mu \Phi ^{\left( t+1 \right)}\left( \mathbf{a}_R \right) \\
&\,\mathrm{s}.\mathrm{t}. \,\,\,0\leqslant a_{n,R}\leqslant 1,  a_{n,T}+a_{n,R}\leqslant 1, \,\forall n,\\
&\,\,\,\,\,\,\,\,\,\,\,    \sum_{n=1}^N{\left( a_{n,T}+a_{n,R} \right)}\leqslant N_{\mathrm{act}},\,\,\eqref{a_r_sensing_constraint},
\end{aligned}
\vspace{-0.3em}\end{equation}
which is ready to be solved by general-purpose convex solvers, where $\Phi ^{( t+1 )}( \mathbf{a}_R ) =\sum_{n=1}^N{[ ( 1-2a_{n,R}^{( t )} ) a_{n,R}+( a_{n,R}^{( t )} ) ^2 ]}$.

Note that the above penalty-involved optimization problems are anticipated to push forward $\{a_{n,T}\}$ and $\{a_{n,R}\}$ to be binary, but not in a strict manner. Thus, the update steps for $\{a_{n,T}\}$ and $\{a_{n,R}\}$ are soft ones. After some iterations of such soft updates, we involve the following progressive and adaptive antenna hardening steps to facilitate binary features of antenna assignment modes, which are shown in Algorithm~\ref{algo_hardening}, where $f_{\mathrm{Tx},n}$, $f_{\mathrm{Rx},n}$, and $f_{\mathrm{Off},n}$ represent binary indicators used to mark whether antenna $n$ is identified as a transmit-hardened, receive-hardened, or inactivated antenna, respectively. Note that the antenna-mode decision thresholds $\tau$ are also increased in a dynamic stepwise manner as the iteration index grows to facilitate the effectiveness of antenna hardening.

In summary, we present the AO-based joint beamforming and antenna assignment algorithm to solve \eqref{sum_rate_maximization}, as in Algorithm~\ref{alg_overall_t}. 
The progressive hardening procedure starts when the iteration index exceeds a prescribed threshold, or earlier if the objective value and the antenna-mode variables have already become sufficiently stable via several continuous iterations. Moreover, we have $\delta_f^{(t+1)}=|R^{(t+1)}-R^{(t)}|/|R^{(t)}|$, $\delta_T^{(t+1)}=\|\mathbf{a}_T^{(t+1)}-\mathbf{a}_T^{(t)}\|_2$, and $\delta_R^{(t+1)}=\|\mathbf{a}_R^{(t+1)}-\mathbf{a}_R^{(t)}\|_2$.

\section{Numerical Results}
\vspace{-0.1cm}
The above frameworks are of general validity across different user and target locations, channel models, and parameter settings, etc. Hence, the following results are provided only as a representative numerical example. We consider an ISAC network with a square coverage area of size $200 \times 200\,\mathrm{m}^2$, defined on the $xz$ plane, where the origin is located at its center. The $y$-axis denotes the vertical direction. A BS is deployed at the center of the coverage area with the BS height $L_{\mathrm{BS}}$, and is equipped with a UPA mounted on the $xy$ plane. The number of UEs is $K=10$. The number of antennas in the UPA in the horizontal and vertical directions are $N_x$ and $N_y$, respectively. Then, let the coordinates of the bottom left point of the BS be $\mathbf{r}_{\mathrm{BS}}=[x_{\mathrm{BS}},y_{\mathrm{BS}},z_{\mathrm{BS}}]^T\in \mathbb{R} ^3$ with $y_{\mathrm{BS}}=L_{\mathrm{BS}}$. Meanwhile, we have $x_{\mathrm{BS}}=z_{\mathrm{BS}}=0$. Then, we derive the position of the $n$-th antenna for the BS as $\mathbf{r}_n=[r_{n,x},r_{n,y},r_{n,z}]^T=[x_{\mathrm{BS}}+\mathrm{mod(}n-1,N_x)\Delta _x,L_{\mathrm{BS}}+\lfloor (n-1)/N_x \rfloor \Delta _y,z_{\mathrm{BS}}]^T\in \mathbb{R} ^3$ by counting the antenna index row-by-row from the bottom left. $\Delta _x$ and $\Delta _y$ denote the horizontal and vertical antenna spacing, respectively. All UEs and the sensing target are randomly distributed in the coverage area. We define the position of the $k$-th UE as $\mathbf{s}_k=[s_{k,x},s_{k,y},s_{k,z}]^T\in \mathbb{R} ^3$ with $s_{k,y}=L_{\mathrm{UE}}$ being the height for each UE. Meanwhile, the position of the sensing target is $\mathbf{s}_0=[s_{0,x},s_{0,y},s_{0,z}]^T\in \mathbb{R} ^3$ with $s_{0,y}=L_{0}$ being the height for the sensing target.

For the communication channel, we have $\mathbf{h}_k=\sqrt{\beta _k}[e^{-j\frac{2\pi}{\lambda}(d_{1k}-d_k)},\dots ,e^{-j\frac{2\pi}{\lambda}(d_{Nk}-d_k)}]^T$, where $\beta _k=\left( \lambda/(4\pi d_k) \right) ^2$ is the free-space channel gain between the UE and BS with $\lambda$ being the wavelength, where $d_k=\left\| \mathbf{r}_{\mathrm{BS}}-\mathbf{s}_k \right\| $ is the distance between the reference point at the BS and UE $k$ and $d_{nk}=\left\| \mathbf{r}_n-\mathbf{s}_k \right\| $ represents the distance between the $n$-th antenna of the BS and UE $k$. For the sensing channel, we have $\mathbf{g}_0=\sqrt{\beta _0}[e^{-j\frac{2\pi}{\lambda}(d_{10}-d_0)},\dots ,e^{-j\frac{2\pi}{\lambda}(d_{N0}-d_0)}]^T$, where $\beta _0=\sqrt{\frac{\lambda ^2}{\left( 4\pi \right) ^3d_{0}^{4}}}$, $d_0=\left\| \mathbf{r}_{\mathrm{BS}}-\mathbf{s}_0 \right\| $, and $d_{n0}=\left\| \mathbf{r}_n-\mathbf{s}_0 \right\| $. For $\mathbf{H}_{\mathrm{SI}}\in \mathbb{C} ^{N\times N}$, we denote its $(i,j)$-th element being $[ \mathbf{H}_{\mathrm{SI}} ] _{ij}=\sqrt{\alpha _{\mathrm{SI}}}e^{j\phi _{ij}}$ with $\phi _{ij}\sim \mathcal{U} \left[ 0,2\pi \right] $ \cite{10158711}. This SI model captures the residual self-interference after efficient SI suppression \cite{6832464}, where a common power parameter captures the residual SI level, and the phase of each element is modeled as random to reflect residual uncertainty and hardware impairments. Unless otherwise stated, we have $L_{\mathrm{BS}}=12.5 \, \mathrm{m}$, $L_{\mathrm{UE}}=L_{0}=1.5 \, \mathrm{m}$, $P_\mathrm{max}=20 \, \mathrm{W}$, $\sigma _{0}^{2}=1\,\mathrm{m}^2$, $\sigma _{k}^{2} =\sigma _{r}^{2}=-80 \, \mathrm{dBm}$, $\alpha_{\mathrm{SI}}=-110\, \mathrm{dB}$, $B=100$, carrier frequency $f_c=3 \, \mathrm{GHz}$, and $\Delta _x=\Delta _y=\lambda/2$. For the AO algorithm, we initialize the communication precoding schemes as the MMSE ones and the sensing precoding scheme as the maximum ratio one. Meanwhile, we initialize the antenna assignment variables from the greedy approach in a softened manner. For an antenna initially assigned to the transmit mode or the receive mode, we set $(a_{n,T},a_{n,R})=(0.6,0.3)$ and $(a_{n,T},a_{n,R})=(0.3,0.6)$, respectively. Besides, we have $\epsilon_{\mathrm{obj}}=10^{-4}$, $\epsilon_T=10^{-3}$, and $\epsilon_R=10^{-3}$.
The simulation results are generated based on the Monte Carlo method with many random realizations of UEs/target locations in the coverage area.

\begin{figure}[t]
\centering
\includegraphics[width=0.8\columnwidth]{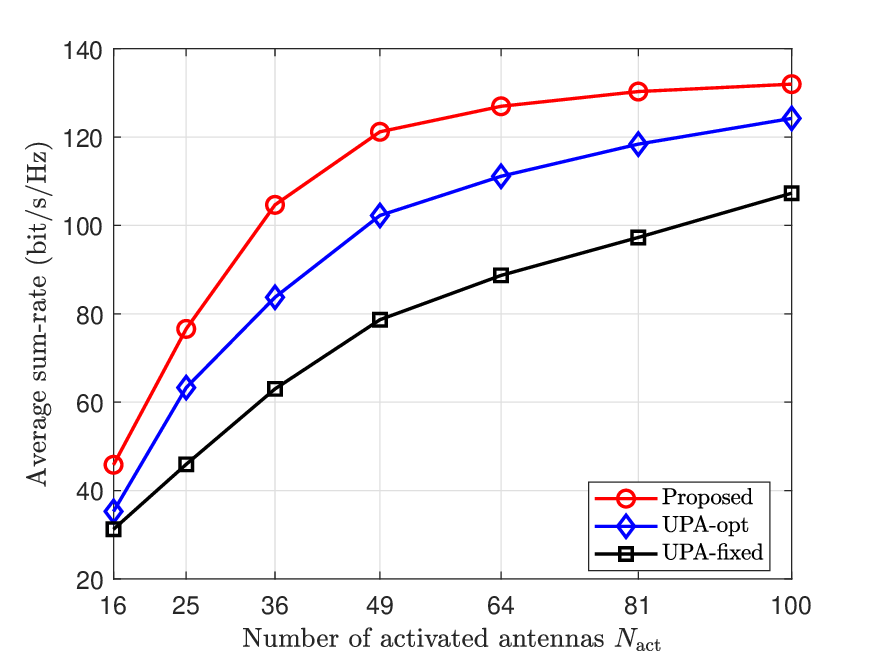}\vspace*{-0.4cm}
\caption{Average sum-rate for different array schemes versus the number of activated antennas $N_{\mathrm{act}}$ with $N_x=20$, $N_y=6$, and $\gamma_0=15\, \mathrm{dB}$. ``Proposed" denotes the proposed non-uniform array scheme.
\label{Fig_Scheme_N_act}}
\vspace{-0.35cm}
\end{figure}

\begin{figure}[t]
\centering
\includegraphics[width=0.8\columnwidth]{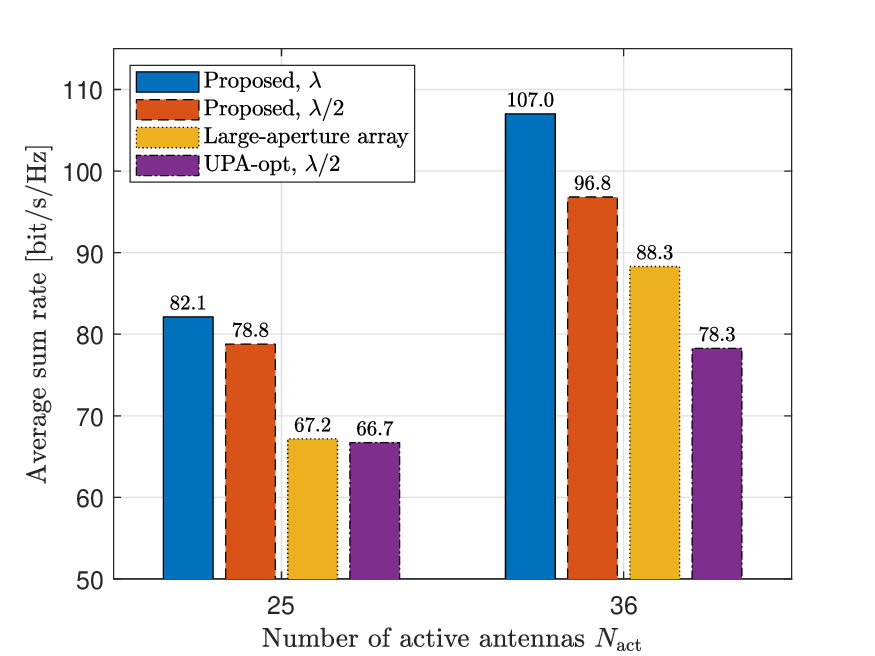}\vspace*{-0.4cm}
\caption{Average sum-rate for different array schemes with values of different antenna spacing $\Delta _x=\Delta _y$, $N_x=12$, $N_y=4$, and $\gamma_0=15\, \mathrm{dB}$. ``UPA-opt, $\lambda/2$"  denotes the baseline using a square UPA with $N_{\mathrm{act}}$ antennas and $\lambda/2$ antenna spacing, optimized following Algorithm~\ref{alg_overall_t}.
\label{Fig_Antenna_Spacing}}
\vspace{-0.1cm}
\end{figure}

\begin{figure}[t]
\centering
\includegraphics[width=0.8\columnwidth]{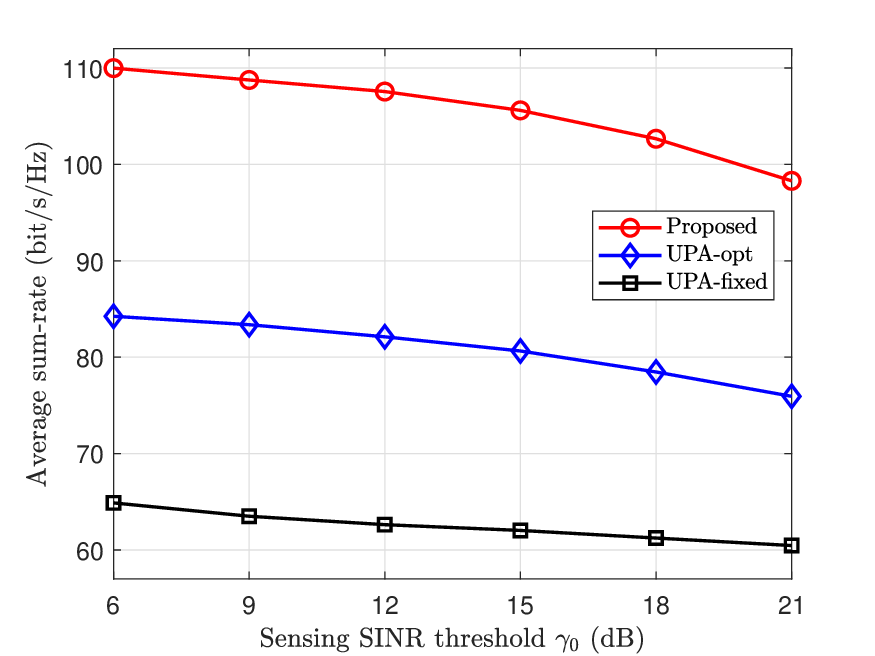}\vspace*{-0.4cm}
\caption{Average sum-rate for different array schemes versus the sensing SINR threshold $\gamma_0$ over $N_x=20$, $N_y=6$, and $N_{\mathrm{act}}=36$. The terminologies of the presented schemes follow the same definitions as in Fig.~\ref{Fig_Scheme_N_act}.
\label{Fig_Scheme_Gamma0}}
\vspace{-0.1cm}
\end{figure}

Fig.~\ref{Fig_Scheme_N_act} shows that the proposed non-uniform array scheme consistently outperforms both the UPA-opt and UPA-fixed baselines.  ``UPA-opt" denotes the baseline using a square UPA with $N_{\mathrm{act}}$ antennas, where the beamforming and antenna assignment variables are optimized following Algorithm~\ref{alg_overall_t}. ``UPA-fixed'' denotes the baseline employing a square UPA with $N_{\mathrm{act}}$ antennas and fixed predefined non-overlapping transmit and receive subarrays, where the left half of the array is assigned to transmission and the right half to reception, and the remaining variables are optimized via Algorithm~\ref{alg_overall_t}. For the same $N_{\mathrm{act}}$, under the same optimization framework, the proposed scheme still achieves a higher sum-rate than the UPA-opt baseline, indicating that the gain mainly stems from the more flexible non-uniform array geometry and the resulting enlarged effective aperture, rather than from the activated antenna budget itself. Moreover, the performance improvement for the proposed scheme compared to the UPA-opt baseline is more pronounced when $N_{\mathrm{act}}$ is small, e.g., $30.0\%$ and $6.2\%$ for $N_{\mathrm{act}}=16$ and $N_{\mathrm{act}}=100$, respectively, suggesting that the benefit of the non-uniform array is particularly significant in the antenna-limited regime, where spatial selection flexibility is more critical. Notably, the proposed scheme with fewer activated antennas can already approach the performance of the UPA-opt baseline with more activated antennas. For instance, the performance gap between the proposed scheme with $N_{\mathrm{act}}=49$ and the UPA-opt with $N_{\mathrm{act}}=100$ is only $2.4\%$ with a $51\%$ reduction in antenna count. These results show that the non-uniform array can efficiently explore potential performance benefits for the activated antennas by offering greater spatial design flexibility than a compact UPA, thereby achieving a target sum-rate with fewer activated antennas.

Fig.~\ref{Fig_Antenna_Spacing} further highlights the impact of array geometry by showing the sum-rate under different antenna spacings. ``Large-aperture array" denotes a rectangular planar array baseline with $N_{\mathrm{act}}$ antennas, embracing the same overall aperture as the $N_x \times N_y$ $\lambda/2$-spaced candidate array. As observed, increasing the antenna spacing from $\lambda/2$ to $\lambda$ improves the sum-rate of the proposed scheme, and this improvement becomes larger when more antennas are activated, e.g., $10.5\%$ improvement for $N_{\mathrm{act}}=36$. This suggests that, under a fixed activated antenna budget, increasing the antenna spacing is a potential way of improving the achievable sum-rate. Besides, the comparison with the large-aperture baseline shows that the gain of the proposed scheme is not merely due to aperture enlargement. Even under the same aperture size, the proposed non-uniform design still provides a non-negligible sum-rate improvement, thereby highlighting the additional benefit of array-geometry flexibility beyond aperture expansion. Moreover, the proposed scheme with $N_{\mathrm{act}}=25$ over different antenna spacings can all outperform the $\lambda/2$-spaced UPA-opt baseline with $N_{\mathrm{act}}=36$, despite using much fewer activated antennas. This demonstrates how we can reduce hardware utilization through properly optimized non-uniform arrays.

Fig.~\ref{Fig_Scheme_Gamma0} shows the average sum-rate of different array schemes versus $\gamma_0$. As expected, the sum-rate of all schemes decreases monotonically with $\gamma_0$, reflecting the fundamental tradeoff between communication performance and sensing requirements. Nevertheless, the proposed scheme consistently achieves the highest sum-rate for all threshold values, followed by the UPA-opt and UPA-fixed baselines. This stable ordering indicates that the advantage of the proposed architecture is not confined to a specific operating point but remains robust under progressively tighter sensing quality-of-service requirements. Moreover, the persistent gap between UPA-opt and UPA-fixed confirms the benefit of jointly optimizing the beamforming and antenna assignment variables, while the further performance gain of the proposed scheme over UPA-opt further highlights the importance of non-uniform array geometry in improving the communication-sensing tradeoff.

\vspace{-0.2cm}
\section{Conclusion}\label{sec:conclusion}
\vspace{-0.1cm}
This paper studied a non-uniform array design for monostatic ISAC systems. By jointly optimizing the ISAC beamforming schemes and triple antenna-mode assignment under sensing, power, and activated-antenna constraints, a sum communication rate maximization problem was formulated and solved via an alternating-optimization-based framework with WMMSE reformulation, continuous relaxation-based penalty involvement, SCA, and progressive antenna hardening. Numerical results demonstrated that the proposed non-uniform array achieves a significantly higher sum-rate than the uniform-array baselines, with the gain being particularly large in the antenna-limited regime. Meanwhile, the proposed design can match, and in some cases outperform, uniform array baselines with substantially fewer activated antennas. These findings confirm that the benefits of non-uniform arrays are fundamentally tied to array geometry, and this flexible geometry-aware design offers a promising alternative to brute-force antenna scaling-based design in future MIMO systems.





\bibliographystyle{IEEEtran}

\bibliography{IEEEabrv,Ref}

\begin{thebibliography}{10}
\providecommand{\url}[1]{#1}
\csname url@samestyle\endcsname
\providecommand{\newblock}{\relax}
\providecommand{\bibinfo}[2]{#2}
\providecommand{\BIBentrySTDinterwordspacing}{\spaceskip=0pt\relax}
\providecommand{\BIBentryALTinterwordstretchfactor}{4}
\providecommand{\BIBentryALTinterwordspacing}{\spaceskip=\fontdimen2\font plus
\BIBentryALTinterwordstretchfactor\fontdimen3\font minus \fontdimen4\font\relax}
\providecommand{\BIBforeignlanguage}[2]{{%
\expandafter\ifx\csname l@#1\endcsname\relax
\typeout{** WARNING: IEEEtran.bst: No hyphenation pattern has been}%
\typeout{** loaded for the language `#1'. Using the pattern for}%
\typeout{** the default language instead.}%
\else
\language=\csname l@#1\endcsname
\fi
#2}}
\providecommand{\BIBdecl}{\relax}
\BIBdecl

\bibitem{ZheSurvey}
Z.~{Wang}, J.~{Zhang} \emph{et~al.}, ``A tutorial on extremely large-scale {MIMO} for {6G}: Fundamentals, signal processing, and applications,'' \emph{IEEE Commun. Surveys Tuts.}, vol.~26, no.~3, pp. 1560--1605, 3rd quarter, 2024.

\bibitem{5595728}
T.~L. Marzetta, ``Noncooperative cellular wireless with unlimited numbers of base station antennas,'' \emph{IEEE Trans. Wireless Commun.}, vol.~9, no.~11, pp. 3590--3600, Nov. 2010.

\bibitem{6736761}
E.~G. Larsson, O.~Edfors \emph{et~al.}, ``Massive {MIMO} for next generation wireless systems,'' \emph{IEEE Commun. Mag.}, vol.~52, no.~2, pp. 186--195, Feb. 2014.

\bibitem{bjornson2026antenna}
E.~Bj{\"o}rnson, A.~Irshad \emph{et~al.}, ``From antenna abundance to antenna intelligence in {6G} gigantic {MIMO} systems,'' \emph{arXiv:2601.08326}, 2026.

\bibitem{10906511}
L.~Zhu, W.~Ma \emph{et~al.}, ``A tutorial on movable antennas for wireless networks,'' \emph{IEEE Commun. Surveys Tuts.}, vol.~28, pp. 3002--3054, 2026.

\bibitem{wang2025flexible}
Z.~Wang, J.~Zhang \emph{et~al.}, ``Flexible {MIMO} for future wireless communications: Which flexibilities are possible?'' \emph{IEEE Wireless Commun.}, vol.~33, no.~1, pp. 181--190, Feb. 2026.

\bibitem{irshad2025pre}
A.~Irshad, A.~Kosasih \emph{et~al.}, ``Pre-optimized irregular arrays versus moveable antennas in multi-user {MIMO} systems,'' \emph{IEEE Wireless Commun. Lett.}, vol.~14, no.~8, pp. 2656--2660, Aug. 2025.

\bibitem{9465145}
I.~Aboumahmoud, A.~Muqaibel \emph{et~al.}, ``A review of sparse sensor arrays for two-dimensional direction-of-arrival estimation,'' \emph{IEEE Access}, vol.~9, pp. 92\,999--93\,017, 2021.

\bibitem{10545312}
X.~Li, Z.~Dong \emph{et~al.}, ``Multi-user modular {XL-MIMO} communications: Near-field beam focusing pattern and user grouping,'' \emph{IEEE Trans. Wireless Commun.}, vol.~23, no.~10, pp. 13\,766--13\,781, Oct. 2024.

\bibitem{9737357}
F.~Liu, Y.~Cui \emph{et~al.}, ``Integrated sensing and communications: Toward dual-functional wireless networks for {6G} and beyond,'' \emph{IEEE J. Sel. Areas Commun.}, vol.~40, no.~6, pp. 1728--1767, Jun. 2022.

\bibitem{11159291}
J.~Zhao, S.~Xue \emph{et~al.}, ``Near-field integrated sensing and communications for secure {UAV} networks,'' \emph{IEEE J. Sel. Areas Commun.}, vol.~44, to appear, 2026.

\bibitem{TWC23CKJ1}
K.~Chen, C.~Qi \emph{et~al.}, ``Simultaneous beam training and target sensing in {ISAC} systems with {RIS},'' \emph{{IEEE} Trans. Wireless Commun.}, vol.~23, no.~4, pp. 2696--2710, Apr. 2024.

\bibitem{10158711}
Z.~He, W.~Xu \emph{et~al.}, ``Full-duplex communication for {ISAC}: Joint beamforming and power optimization,'' \emph{IEEE J. Sel. Areas Commun.}, vol.~41, no.~9, pp. 2920--2936, Sep. 2023.

\bibitem{9724187}
Z.~Xiao and Y.~Zeng, ``Waveform design and performance analysis for full-duplex integrated sensing and communication,'' \emph{IEEE J. Sel. Areas Commun.}, vol.~40, no.~6, pp. 1823--1837, Jun. 2022.

\bibitem{10159012}
Z.~Liu, S.~Aditya \emph{et~al.}, ``Joint transmit and receive beamforming design in full-duplex integrated sensing and communications,'' \emph{IEEE J. Sel. Areas Commun.}, vol.~41, no.~9, pp. 2907--2919, Sep. 2023.

\bibitem{10810291}
R.~Liu, M.~Li \emph{et~al.}, ``{DOA} estimation-oriented joint array partitioning and beamforming designs for {ISAC} systems,'' \emph{IEEE Trans. Wireless Commun.}, vol.~24, no.~3, pp. 2052--2066, Mar. 2025.

\bibitem{8187178}
E.~{Bj{\"o}rnson}, J.~{Hoydis}, and L.~{Sanguinetti}, \emph{Massive MIMO Networks: Spectral, Energy, and Hardware Efficiency}.\hskip 1em plus 0.5em minus 0.4em\relax now Publishers Inc, 2017.

\bibitem{5756489}
Q.~Shi, M.~Razaviyayn \emph{et~al.}, ``An iteratively weighted {MMSE} approach to distributed sum-utility maximization for a {MIMO} interfering broadcast channel,'' \emph{IEEE Trans. Signal Process.}, vol.~59, no.~9, pp. 4331--4340, Apr. 2011.

\bibitem{6832464}
A.~Sabharwal, P.~Schniter \emph{et~al.}, ``In-band full-duplex wireless: Challenges and opportunities,'' \emph{IEEE J. Sel. Areas Commun.}, vol.~32, no.~9, pp. 1637--1652, Sep. 2014.

\end{thebibliography}

\end{document}